%% file: Main.tex
\newcommand{\CLASSINPUTtoptextmargin}{19mm}%
\newcommand{\CLASSINPUTbottomtextmargin}{43mm}%
\newcommand{\CLASSINPUTinnersidemargin}{12.9mm}%
\newcommand{\CLASSINPUToutersidemargin}{12.9mm}%
\documentclass[conference,10pt,a4paper]{IEEEtran}
\usepackage{amsmath}
\usepackage{times}
\usepackage{graphicx}
\usepackage{multirow}
\usepackage[none]{hyphenat}
\usepackage{float}
\usepackage{subfig}
\usepackage{booktabs}
\usepackage{eso-pic}

\input{EuMW_modify_IEEEtran_18b_CTAN_V4}

\begin{document}
\raggedbottom
%
%
%
\title{Inverse Design of Compact and Wideband Inverted Doherty Power Amplifiers Using Deep Learning}
%
%
\author{%
\IEEEauthorblockN{%
Han~Zhou\textsuperscript{\protect\#}, Haojie~Chang\textsuperscript{\protect\#}, David Widén\textsuperscript{\protect\&}, Christian~Fager\textsuperscript{\protect\&}
}

\IEEEauthorblockA{%
\textsuperscript{\protect\#}Faculty of Information Technology and Communication Sciences, Tampere University, Finland\\
\textsuperscript{\protect\&}Department of Microtechnology and Nanoscience, Chalmers University of Technology, Sweden\\
han.zhou@tuni.fi
}
}
%

\AddToShipoutPictureFG*{%
  \AtPageUpperLeft{%
    \hspace{0.63in}%
    \raisebox{-0.32in}{%
      \parbox{\dimexpr\paperwidth-1.26in\relax}{%
        \centering\footnotesize\itshape
        This is the author accepted version of a paper accepted for presentation at the
        56th European Microwave Conference (EuMC 2026).
        \copyright~2026 IEEE. The final published version will be available in IEEE Xplore.
      }%
    }%
  }%
}
\maketitle
%
%
\begin{abstract}
This paper presents a deep learning‑assisted methodology for the inverse synthesis of a compact, wideband inverted Doherty power amplifier (PA). Convolutional neural networks (CNNs) and genetic algorithms (GAs) are jointly employed to generate pixelated Doherty combiner networks that integrate load modulation, impedance matching, power combining, and phase compensation into a single structure. As a proof of concept, we design and fabricate a GaN HEMT Doherty PA with a pixelated output combiner. The prototype achieves a measured peak drain efficiency of 51\%--63\% and a 6-dB back‑off efficiency of 48\%--54\% over 1.9--2.5~GHz. Within the same frequency range, the measured output power is 44$\pm$0.3~dBm. Furthermore, with digital predistortion (DPD) applied, the prototype circuit demonstrates an adjacent channel leakage ratio (ACLR) better than -53.2~dBc.

\end{abstract}
%

\section{Introduction}
\label{sec:Introduction}

As data throughput continues to increase dramatically and more complex modulation schemes are adopted, power amplifiers (PAs) must maintain high efficiency at significantly reduced output power levels while also operating over wider frequency ranges. The Doherty PA remains one of the most widely deployed architectures because it provides substantial back-off efficiency enhancement with a simple circuit implementation. This advantage is especially notable when compared with more recently proposed load‑modulated PA architectures \cite{LMBA1, DEPA1, SLMBA2, NOLMPA, CLMA2}. However, designing the Doherty combiner is highly challenging because the main and auxiliary amplifiers interact directly through this network, and this strong interaction also limits the achievable bandwidth.

\begin{figure}[t!]
    \centering
    \subfloat[]{
        \includegraphics[width=0.85\linewidth]{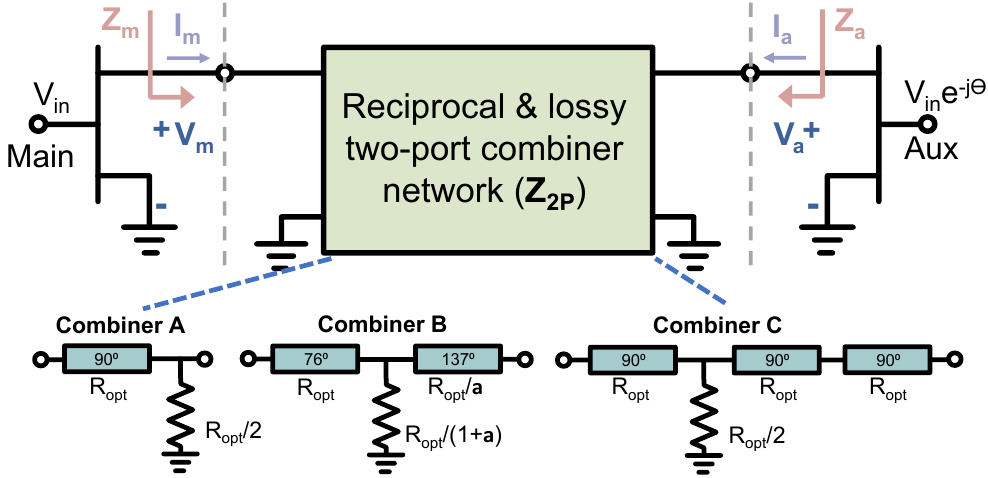}
    }
    \vspace{1mm}
    \subfloat[]{
        \includegraphics[width=0.65\linewidth]{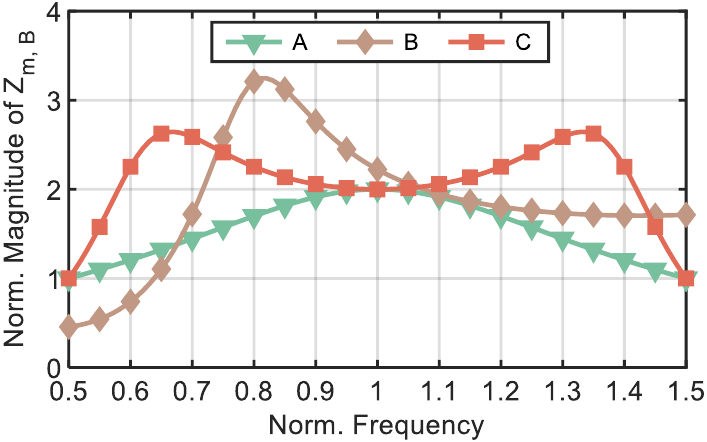}
    }
    \caption{(a) Synthesis of the Doherty combiner with three implementations: Combiner~A (conventional), Combiner~B (black‑box), and Combiner~C (inverted). (b) Magnitude of the impedances presented to the main and auxiliary current‑source planes at back‑off for Combiners~A–C~\cite{zhou2023phd}.}
    \label{fig:}
\end{figure}
Recently, inverse design techniques based on pixelated circuit layouts and deep learning have been introduced for RF circuit design~\cite{DL_PA_JSSC}. These methods explore arbitrary electromagnetic (EM) layouts and can uncover non‑intuitive solutions that are more capable of locating globally optimal designs. As a result, they have been successfully applied to a variety of PA implementations across different frequency bands and processes~\cite{DL_PA_JSSC, AI_HZ, DL_6GPA}. In~\cite{TCASI_AIHan}, we extended this approach to the Doherty PA by treating the Doherty combiner as a lossy and reciprocal two‑port network, as illustrated in Fig. 1(a). We first combined the black‑box Doherty design method with deep learning to synthesize a pixelated three‑port Doherty combiner directly from load‑pull data. The black‑box Doherty method is advantageous because it provides an extended dynamic range when symmetric transistors are used. However, as shown in Fig. 1, the Combiner~B synthesized using this approach exhibits degraded bandwidth performance. To address this limitation, we proposed a dual‑state impedance synthesis Doherty design methodology in~\cite{IMS_AIHan} to realize the conventional Doherty PA (Combiner~A) with slightly improved bandwidth. Nevertheless, the bandwidth of the conventional Doherty PA remains limited. The inverted Doherty PA, originally proposed in~\cite{IDPA1, DohertyJames} and corresponding to Combiner~C, offers significantly greater bandwidth potential.

\begin{figure*} [t!]
    \centering       
    \includegraphics[width=0.85\linewidth]{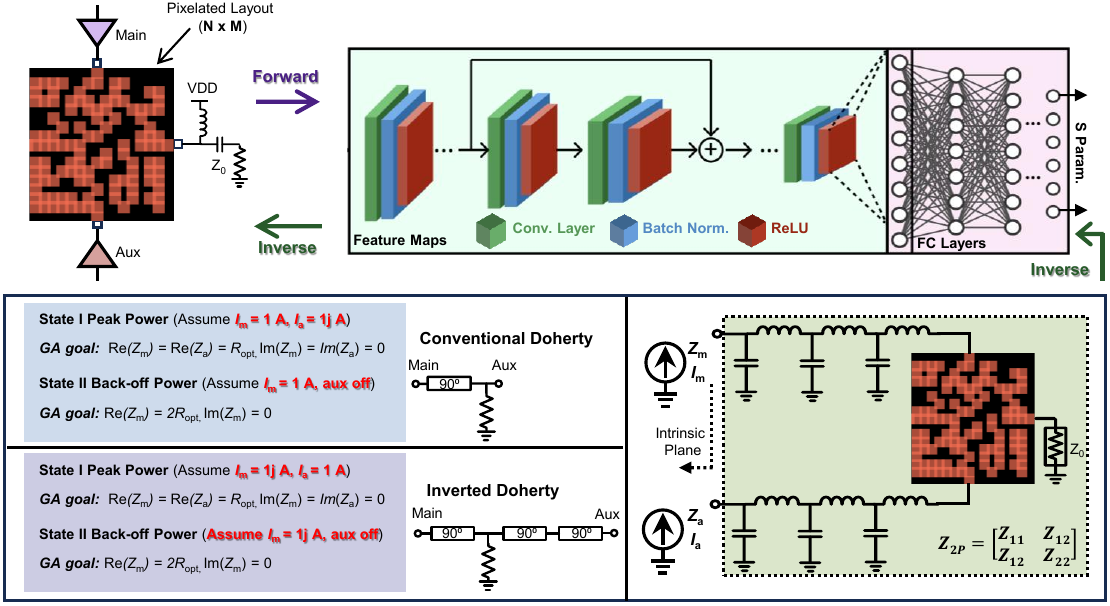}
    \caption{Inverse‑design workflow for pixelated conventional and inverted Doherty combiners using deep convolutional neural networks (CNNs) and a genetic algorithm integrated with the dual‑state impedance‑synthesis method proposed in~\cite{IMS_AIHan}.}
    \label{fig.2}
\end{figure*}
In this paper, we propose the first inverted Doherty PA design driven by deep learning. We leverage the dual‑state impedance synthesis approach and train deep convolutional neural networks (CNN) as a surrogate model to replace full‑wave EM simulation, allowing us to rapidly capture the nonlinear relationships between pixelated EM structures and S‑parameters. Together with a genetic algorithm (GA), we have successfully designed and validated a GaN HEMT Doherty PA prototype that use compact pixelated combiners and achieve excellent measured performance over a wide bandwidth.

\section{Theory}
\subsection{Inverted Doherty Output Combiner Synthesis with Dual-State Impedance Approach}
As shown in Fig. 2, the overall Doherty combiner can be treated as a reciprocal and lossy two‑port network (denoted as \textbf{A}\textsubscript{2p}) when the load port is terminated with 50~$\Omega$. The transistor parasitics and the equivalent packaged circuit can be extracted from either simulation or measurement~\cite{AI_HZ}, and their ABCD matrix is denoted as \textbf{A}\textsubscript{par}. Together with the pixelated Doherty output network \textbf{A}\textsubscript{pixel}, the ABCD matrix of the overall two‑port network presented to the Doherty PA transistor’s current‑source plane can be expressed as~\cite{IMS_AIHan}
\begin{equation}
\mathbf{A_{\mathrm{2p}}} = \mathbf{A_{\mathrm{par}}A_{\mathrm{pixel}}A_{\mathrm{par}}}
\end{equation}
It is then straightforward to convert the \textbf{A}\textsubscript{2p} into the impedance matrix \textbf{Z}\textsubscript{2p}. The matrix \textbf{Z}\textsubscript{2p} characterizes the interaction between the main and auxiliary amplifiers, and the following relation holds

\begin{equation}
\begin{bmatrix}
V_{\text{m}} \\
V_{\text{a}}
\end{bmatrix}
=
Z_{\text{2p}}
\begin{bmatrix}
I_{\text{m}} \\
I_{\text{a}}
\end{bmatrix}
\end{equation}

Since the impedance matrix Z\textsubscript{2p} depends only on the pixelated combiner (the parasitic and packaged elements are already known), we can easily set up a simple AC simulation to determine the impedances seen at the main and auxiliary current‑source planes. Note that we assume a symmetrical Doherty configuration, which means that at peak power the condition $|I_{\mathrm{a}}| = |I_{\mathrm{m}}|$ holds, and that at back‑off power the auxiliary amplifier is open for both the conventional and inverted Doherty cases.

\begin{itemize}
  \item \textbf{Conventional Doherty PA:}  
  Since $\angle(I_{\mathrm{a}}, I_{\mathrm{m}}) = 90^\circ$, at peak power, we assume  
  $I_{\mathrm{m}} = 1\,\mathrm{A}$ and $I_{\mathrm{a}} = 1j\,\mathrm{A}$.  
  At back-off power, we set $I_{\mathrm{m}} = 1\,\mathrm{A}$.

  \item \textbf{Inverted Doherty PA:}  
  Since $\angle(I_{\mathrm{a}}, I_{\mathrm{m}}) = -90^\circ$, at peak power, we assume  
  $I_{\mathrm{m}} = 1j\,\mathrm{A}$ and $I_{\mathrm{a}} = 1\,\mathrm{A}$.  
  At back-off power, we set $I_{\mathrm{m}} = 1j\,\mathrm{A}$.
\end{itemize}

\begin{figure} [t!]
    \centering    
    \includegraphics[width=0.75\columnwidth]{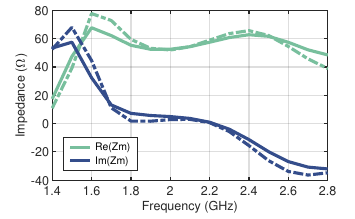}
    \caption{Comparison of CNN‑predicted (dashed) and EM‑simulated (solid) impedance at the main current‑source planes of the synthesized inverted Doherty combiner at back‑off power.}
    \label{fig.6}
\end{figure}

\subsection{Deep Convolutional Neural Network and Genetic Algorithm for Pixelated Doherty Combiner Generation}
We encode each planar circuit layout as a binary $15{\times}15$~grid, where "1" denotes metal and "0" represents empty space. Each pixel measures $1.2~\mathrm{\times}~1.2~\mathrm{mm}$ and a 40\% metal overlap is applied to guarantee reliable diagonal connectivity. To train the deep CNN, we generated a dataset of 5000 layouts together with their EM‑simulated S‑parameters. The layouts follow a normally distributed metal density (mean~50\%, standard deviation~15\%). Python scripts automatically controlled ADS Momentum to run the EM simulations, and we expanded the dataset by four times through flipping and rotation for augmentation. Figure~2 illustrates the deep~CNN structure with residual connections. The model takes the binary layout matrix as input and predicts the corresponding S‑parameters. The network contains twelve convolutional layers followed by five fully connected (FC) layers. Each convolution stage uses batch normalization and a leaky‑ReLU activation. After feature extraction, the output feature maps are flattened before the FC layers. Dropout regularization is applied during training to reduce overfitting. 

After training, the CNN surrogate is paired with a GA to explore the layout space efficiently. Tournament selection and an injection‑type mutation introduce fresh candidate matrices to maintain population diversity. The GA optimizes the combiner layout so that, over 1.8--2.6~GHz:
\begin{figure} [t!]
    \centering    
    \includegraphics[width=0.6\columnwidth]{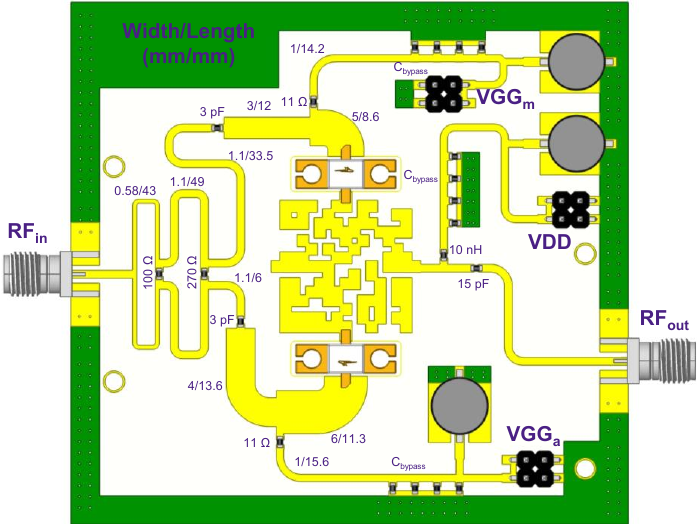}
    \caption{Circuit schematic of the proposed Inverted Doherty PA.}
    \label{fig.6}
\end{figure}
\begin{figure} [t!]
    \centering    
    \includegraphics[width=0.6\columnwidth]{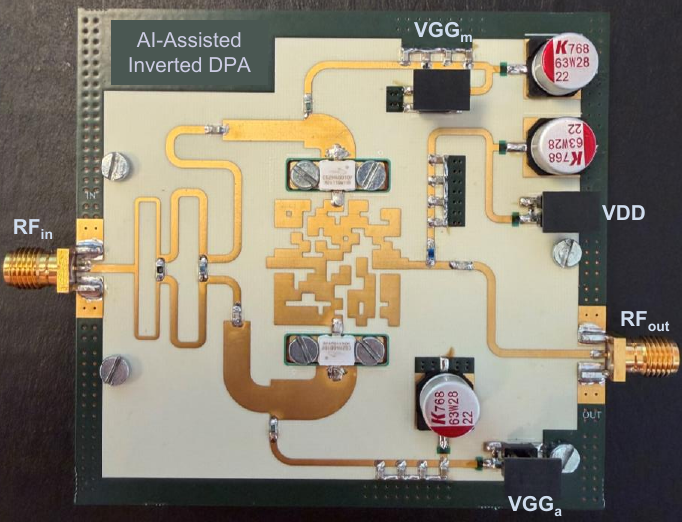}
    \caption{The fabricated prototype circuit with dimensions of $78~\mathrm{mm}~\mathrm{\times}~72~\mathrm{mm}$.}
    \label{fig.6}
\end{figure}
\begin{itemize}
    \item $\Re(Z_{\mathrm{m,peak}})$ \& $\Re(Z_{\mathrm{a,peak}})$ stay within $\pm 10\%$ of $R_{\text{opt}}$,
    \item $\Re(Z_{\mathrm{m,bo}})$ remains within $\pm 10\%$ of $2R_{\text{opt}}$,
\end{itemize}
while minimizing the imaginary components.
\begin{figure} [t!]
    \centering    
    \includegraphics[width=0.8\columnwidth]{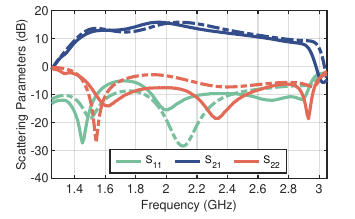}
    \caption{Measured small-signal results of the fabricated prototype circuit.}
    \label{fig.7}
\end{figure}
\begin{figure} [t!]
    \centering        
    \includegraphics[width=0.8\columnwidth]{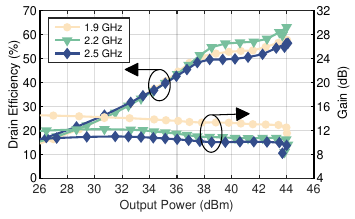}
    \caption{Measured drain efficiency and gain of the fabricated prototype circuit as functions of output power over the $1.9-2.5~\mathrm{GHz}$ frequency range.}
    \label{fig.8}
\end{figure}

\begin{figure} [t!]
    \centering        
    \includegraphics[width=0.8\columnwidth]{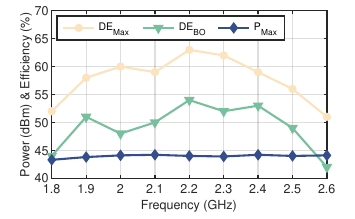}
    \caption{Measured saturated output power, peak drain efficiency, and 6‑dB back‑off drain efficiency of the fabricated prototype versus frequency.}
    \label{fig.9}
\end{figure}

As illustrated in Fig.~3, the synthesized inverted      Doherty combiner shows good agreement between the CNN‑predicted impedances and full‑wave EM results at back‑off. The employed transistor exhibits an optimal load of $R_{\mathrm{opt}}=30~\Omega$, consistent with the target impedance goals.

\section{Prototype Design and Measurement Results}
Fig.~4 presents the complete schematic of the proposed inverted Doherty PA. The prototype is implemented on a 20‑mil Rogers 4350B substrate and employs two 10‑W packaged GaN~HEMT devices from Macom serving as the main and auxiliary transistors. As shown in Fig.~5, the fabricated circuit occupies a compact footprint of $78~\mathrm{mm}~\mathrm{\times}~72~\mathrm{mm}$. We conducted small‑signal characterization, large‑signal continuous‑wave (CW) measurements, and modulated‑signal testing to evaluate the performance of the prototype. During all measurements, the main amplifier was biased with a gate voltage of $-2.45~\mathrm{V}$, yielding a quiescent current of approximately $20~\mathrm{mA}$. The auxiliary amplifier gate was biased at $-7~\mathrm{V}$, and the drain bias voltage for both devices was maintained at $28~\mathrm{V}$. 

\begin{figure} [t!]
    \centering        
    \includegraphics[width=0.7\columnwidth]{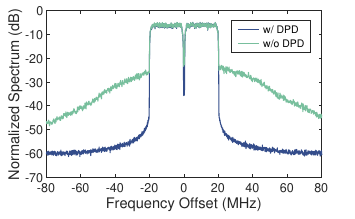}
    \caption{Normalized output spectrum of the fabricated prototype with a $40$-MHz, $7$-dB PAPR OFDM signal at $2.2$~GHz before and after DPD.}
    \label{fig.9}
\end{figure}

The small‑signal measurement results are presented in Fig.~6. The measured response shows good agreement with the EM simulations, with only a slight frequency shift observed at the upper end of the band. The prototype exhibits a small‑signal gain ($S_{21}$) exceeding $10~\mathrm{dB}$ across the $1.8-2.6~\mathrm{GHz}$ frequency range.

The CW measurement results are summarized in Fig.~7 and~8. The measured drain efficiency and gain are plotted as functions of the output power across the $1.9-2.5~\mathrm{GHz}$ frequency range. Clear Doherty‑type efficiency enhancement behavior is observed throughout the design band. Fig.~\ref{fig.9} further consolidates the performance across frequency. Over the $1.9-2.5~\mathrm{GHz}$ bandwidth, the prototype delivers a peak output power of $44 \pm 0.3~\mathrm{dBm}$. The corresponding peak drain efficiency is within $51\%-63\%$ while the efficiency at 6‑dB back‑off remains $48\%-54\%$. The prototype was further evaluated using a 40‑MHz OFDM signal with a 7‑dB PAPR. As shown in Fig.~9, the adjacent channel leakage ratio (ACLR) improves from $-28.6$ to $-53.2$~dBc after applying DPD. Table~\ref{tab.1} compares the fabricated PA with recent load‑modulated PA designs. The proposed prototype achieves high efficiency and wide bandwidth within a compact footprint, demonstrating the effectiveness of the deep learning‑based design methodology.

\begin{table}[t!]
    \centering
    \caption{Summary Of State-of-the-Art Load-Modulated PAs.}
    \begin{tabular}{ c c c c c c c }
    \toprule  
    \multirow{2}{*}{ Ref.} & \multirow{2}{*}{Arch.} & Freq & $\eta$\textsubscript{SAT} & $\eta$\textsubscript{BO-6dB} & P\textsubscript{SAT} & ACLR\\
    & & (GHz) & ($\%$) & ($\%$) & (dBm) & (dBc) \\

    \midrule
    \multirow{1}{*}{ \cite{SLMBA2}'20} & SLMBA & \multirow{1}{*}{3.0--3.5} & 60--74 & 50--64 &  42.3 & \multirow{1}{*}{-46.7} \\
    \midrule
    \multirow{1}{*}{ \cite{OLMBA}'25} & OLMBA & \multirow{1}{*}{0.8--1.6} & 63--66& 48--57 & 46& \multirow{1}{*}{-45.0} \\
    \midrule
    \multirow{1}{*}{ \cite{Table2}'23} & 2-DPA& \multirow{1}{*}{3.0--3.5} & 51--55 & 43--45 &  43.0 & \multirow{1}{*}{-49.1} \\
    
    \midrule
    \multirow{1}{*}{ \cite{Table1}'23} & 2-DPA& \multirow{1}{*}{3.3--3.9} & 48--53 & 34--45 &  45.6 & \multirow{1}{*}{N.A.} \\
 
    \midrule
    \multirow{1}{*}{\textbf{This Work}} & \textbf{DPA 1} & 1.9--2.5 & 51--63 & 48--54 & 44.0 & -53.2 \\
 
    \bottomrule    
    \end{tabular}\\
    \label{tab.1}
\end{table}
\section{Conclusion}
This paper introduces a deep learning–driven methodology for designing an inverted Doherty PA with a pixelated output combiner. By combining a deep CNN surrogate model with a genetic algorithm, the proposed approach enables efficient inverse synthesis of a compact, wideband Doherty output network. A $1.9–2.5~\mathrm{GHz}$ prototype validates the method, achieving $51\%-63\%$ peak efficiency, $48\%-54\%$ back‑off efficiency, and a saturated power of $44\pm0.3~\mathrm{dBm}$. These results demonstrate the strong potential of deep learning–based inverse synthesis with pixelated EM sturctuers in advancing wideband Doherty PA design.
    
\section*{Acknowledgment}
This research was supported by VINNOVA Grant 2024-02531, MULTIRACS, through the Eureka CELTIC Framework. The authors further acknowledge the TIMES group (Tampere Integrated MicroElectronics and Systems), Tampere University. We also thank MACOM for providing the GaN HEMT transistors used in this work.


\bibliographystyle{IEEEtran}

\bibliography{IEEEabrv,mybibfile}

\end{document}

%% file: EuMW_modify_IEEEtran_18b_CTAN_V4.tex
\makeatletter

\def\@maketitle{\newpage
\bgroup\par\addvspace{0.5\baselineskip}\centering%
\ifCLASSOPTIONtechnote
   {\bfseries\large\@IEEEcompsoconly{\sffamily}\@title\par}\vskip 1.3em{\lineskip .5em\@IEEEcompsoconly{\sffamily}\@author
   \@IEEEspecialpapernotice\par{\@IEEEcompsoconly{\vskip 1.5em\relax
   \@IEEEtitleabstractindextextbox{\@IEEEtitleabstractindextext}\par
   \hfill\@IEEEcompsocdiamondline\hfill\hbox{}\par}}}\relax
\else
   \vskip0.2em{\EuMWtitlesize\ifCLASSOPTIONtransmag\bfseries\LARGE\fi\@IEEEcompsoconly{\sffamily}\@IEEEcompsocconfonly{\normalfont\normalsize\vskip 2\@IEEEnormalsizeunitybaselineskip
   \bfseries\Large}\@title\par}\vskip1.0em\par
   \ifCLASSOPTIONconference%
      {\@IEEEspecialpapernotice\mbox{}\vskip\@IEEEauthorblockconfadjspace%
       \mbox{}\hfill\begin{@IEEEauthorhalign}\@author\end{@IEEEauthorhalign}\hfill\mbox{}\par}\relax
   \else
      \ifCLASSOPTIONpeerreviewca
         {\@IEEEcompsoconly{\sffamily}\@IEEEspecialpapernotice\mbox{}\vskip\@IEEEauthorblockconfadjspace%
          \mbox{}\hfill\begin{@IEEEauthorhalign}\@author\end{@IEEEauthorhalign}\hfill\mbox{}\par
          {\@IEEEcompsoconly{\vskip 1.5em\relax
           \@IEEEtitleabstractindextextbox{\@IEEEtitleabstractindextext}\par\hfill
           \@IEEEcompsocdiamondline\hfill\hbox{}\par}}}\relax
      \else
         \ifCLASSOPTIONtransmag
           {\@IEEEspecialpapernotice\mbox{}\vskip\@IEEEauthorblockconfadjspace%
            \mbox{}\hfill\begin{@IEEEauthorhalign}\@author\end{@IEEEauthorhalign}\hfill\mbox{}\par
           {\vspace{0.5\baselineskip}\relax\@IEEEtitleabstractindextextbox{\@IEEEtitleabstractindextext}\vspace{-1\baselineskip}\par}}\relax
         \else
           {\lineskip.5em\@IEEEcompsoconly{\sffamily}\sublargesize\@author\@IEEEspecialpapernotice\par
           {\@IEEEcompsoconly{\vskip 1.5em\relax
            \@IEEEtitleabstractindextextbox{\@IEEEtitleabstractindextext}\par\hfill
            \@IEEEcompsocdiamondline\hfill\hbox{}\par}}}\relax
         \fi
      \fi
   \fi
\fi\par\addvspace{0.0\baselineskip}\egroup}

\def\EuMWtitlesize{\@setfontsize{\EuMWtitlesize}{24}{24pt}}
\def\EuMWauthorsize{\@setfontsize{\EuMWauthorsize}{11}{11pt}}
\def\EuMWaffilsize{\@setfontsize{\EuMWaffilsize}{10}{10pt}}
\def\EuMWcaptionsize{\@setfontsize{\EuMWcaptionsize}{9}{10pt}}
\def\EuMWbibsize{\@setfontsize{\EuMWbibsize}{8}{10pt}}

\def\@IEEEauthorblockNstyle{\EuMWauthorsize\@IEEEcompsocnotconfonly{\sffamily}\@IEEEcompsocconfonly{\large}}
\def\@IEEEauthorblockAstyle{\EuMWaffilsize\@IEEEcompsocnotconfonly{\sffamily}\@IEEEcompsocconfonly{\itshape}\@IEEEcompsocconfonly{\large}}
\def\@IEEEauthordefaulttextstyle{\EuMWauthorsize\@IEEEcompsocnotconfonly{\sffamily}\sublargesize}

\def\thebibliography#1{\section*{\refname}%
    \addcontentsline{toc}{section}{\refname}%
    \EuMWbibsize\@IEEEcompsocconfonly{\small}\vskip 0.3\baselineskip plus 0.1\baselineskip minus 0.1\baselineskip
    \list{\@biblabel{\@arabic\c@enumiv}}%
    {\settowidth\labelwidth{\@biblabel{#1}}%
    \leftmargin\labelwidth
    \advance\leftmargin\labelsep\relax
    \itemsep \IEEEbibitemsep\relax
    \usecounter{enumiv}%
    \let\p@enumiv\@empty
    \renewcommand\theenumiv{\@arabic\c@enumiv}}%
    \let\@IEEElatexbibitem\bibitem%
    \def\bibitem{\@IEEEbibitemprefix\@IEEElatexbibitem}%
\def\newblock{\hskip .11em plus .33em minus .07em}%
\ifCLASSOPTIONtechnote\sloppy\clubpenalty4000\widowpenalty4000\interlinepenalty100%
\else\sloppy\clubpenalty4000\widowpenalty4000\interlinepenalty500\fi%
    \sfcode`\.=1000\relax}

\long\def\@makecaption#1#2{%
\ifx\@captype\@IEEEtablestring%
\par\@IEEEtabletopskipstrut
\else
\@IEEEfigurecaptionsepspace
\fi
\setbox\@tempboxa\hbox{\normalfont\footnotesize {#1.}\nobreakspace\nobreakspace #2}%
\ifdim \wd\@tempboxa >\hsize%
\setbox\@tempboxa\hbox{\normalfont\footnotesize {#1.}\nobreakspace\nobreakspace}%
\parbox[t]{\hsize}{\normalfont\footnotesize\noindent\unhbox\@tempboxa#2}%
\else
\ifCLASSOPTIONconference \hbox to\hsize{\normalfont\footnotesize\hfil\box\@tempboxa\hfil}%
\else \hbox to\hsize{\normalfont\footnotesize\box\@tempboxa\hfil}%
\fi\fi
\ifx\@captype\@IEEEtablestring%
\@IEEEtablecaptionsepspace
\else
\fi}

\newlength\tablecaptiontotableskip
\newlength\figuretocaptionskip
\def\@IEEEfigurecaptionsepspace{\vskip\figuretocaptionskip\relax}%
\def\@IEEEtablecaptionsepspace{\vskip\tablecaptiontotableskip\relax}%

\def\abstract{\normalfont%
\@IEEEabskeysecsize\bfseries\textit{\abstractname}\,\bfseries\textit{---}\,%
\@IEEEgobbleleadPARNLSP}%

\def\IEEEkeywords{\normalfont%
\@IEEEabskeysecsize\bfseries\textit{\IEEEkeywordsname}\,\bfseries\textit{---}\,%
\@IEEEgobbleleadPARNLSP}%
\def\endIEEEkeywords{\relax\vspace{0.67ex}%
\par\if@twocolumn\else\endquotation\fi%
\normalsize\normalfont}%

\def\@IEEEauthorblockNtopspace{0ex}
\def\@IEEEauthorblockAtopspace{1mm}
\def\tablename{Table}
\def\thetable{\arabic{table}}
\def\IEEEkeywordsname{Keywords}
\def\subsubsection{\@startsection{subsubsection}{3}{\z@}{1.5ex plus 1.5ex minus 0.5ex}%
{0.7ex plus .5ex minus 0ex}{\normalfont\normalsize\itshape}}%
\newlength{\CPheadmatchindent}%
\def\@seccntformat#1{\hbox to\CPheadmatchindent{\csname the#1dis\endcsname}\hskip 0.1em \relax}
\IEEEilabelindentA \parindent
\IEEEilabelindent \IEEEilabelindentA
\IEEEelabelindent \parindent
\IEEEdlabelindent \parindent
\IEEElabelindent \parindent
\makeatother